\documentstyle[aps,prl]{revtex}
\input psfig
\tightenlines
\begin{document}
\newcommand{\vect}[1]{ {\bf #1} }
\newcommand{\vectg}[1]{\mbox{\boldmath $#1$}}
\title{Towards a two-fluid picture of intermittency in shell models
of turbulence}
\author{Jean-Louis Gilson and Thierry Dombre}
\address{Centre de Recherches sur les Tr\`es Basses Temp\'eratures-CNRS,
 Laboratoire conventionn\'e avec l'Universit\'e Joseph Fourier, BP166, 
38042 Grenoble Cedex~9, France}
\date{30 May 1997}
\maketitle
\begin{abstract}
Intermittency in the Gledzer-Okhitani-Yamada (GOY) model of turbulence
is explained in terms of collisions of coherent soliton-like structures 
with a random background issuing from the desintegration of their
predecessors. This two-fluid picture is substantiated by the elucidation of
local dynamical mechanisms leading to anomalous growth of coherent
structures, their detection in true signals involving forcing and 
dissipation, and an investigation of their statistics.

\end{abstract}

From a theoretical point of view, one of the most challenging features of
fully developed turbulence is an interplay between random almost
Gaussian background and coherent ordered structures responsible
for deviations from Gaussian statistics. This duality explains in particular
why field theoretical methods meet great difficulties in capturing
intermittency effects directly from Navier-Stokes equations, even if
considerable progress has been made recently in the related problem of random advection \cite{RA,IN}. Although
coherent structures were visualized as sheets or tubes of vorticity \cite
{FI} in the case of 3D-incompressible turbulence, little is yet known about the way they form, their degree of stability and as a consequence their 
statistical relevance.\\

Addressing such issues in the simpler context of the so-called shell models
of turbulence may help one to figure out new mechanisms of intermittency, possibly at work in the complete Navier-Stokes dynamics. 
It was first realized by Siggia \cite{S78} that the 
one-dimensional character of these models favours
the formation of coherent soliton-like pulses, whose amplitude
grows in a self-similar way, as they move from large to small scales.
Many years later, Parisi \cite{P90}, in an unpublished work, envisioned the
 turbulent medium forming in these systems as a gas of
interacting ``solitons'', with a continuous spectrum of scaling exponents
leading to mutifractality. However, we have shown recently \cite {DG97}
 that genuine self-similar solutions of the equations of motion in the
 inertial range display an unique scaling exponent (to be denoted below as
$z_{0}$), provided they are localized in k-space. When $z_{0}$
departs enough from the Komogorov value 2/3, the
scaling of large deviations observed in full simulations of the
corresponding shell model (including energy injection at large
scales and dissipation at small ones) is well accounted for by the
properties of these ideal objects. But the resulting statistics, almost
unifractal, is very different from realistic turbulence.

In contrast to the previous situation, $z_{0}$ turns out to be quite close to
 2/3 in the case of the Gledzer-Okhitani-Yamada (GOY) model \cite{GOY}, 
in the range of parameters where it reproduces very well the 
multiscaling properties of real turbulent flow \cite{LS97}. 
Taken literally, this result
would suggest weak intermittency effects unless the hitherto
ignored dressing of coherent structures by their interaction with the rest
of the flow helps to produce more singular fluctuations.
Acceleration of time scales downward
the cascade, which is a salient feature of Navier-Stokes dynamics well
captured by shell models, makes the collision between two coherent
structures very unlikely. We are thus led rather naturally to a two-fluid
picture, where coherent structures form in and propagate into a 
featureless random background. Somewhat paradoxically, the good transfer
properties of the GOY model turn out to be a crucial ingredient in the story,
allowing 
the persistence of a rich activity on all scales between the passage of two
 intense events. Randomness of the background comes primarily from the
 large separation of time scales between the integral and dissipative scales.
In this Letter, we present various facts supporting
 this new physical picture of intermittency.\\

Equations of motion for the GOY model in the inertial range 
may be cast in the form~:
\begin{equation}
\frac{db_{n}}{dt}  =  Q^{2}(1-\epsilon)b^{*}_{n-2}b^{*}_{n-1} 
+ \epsilon b^{*}_{n-1}b^{*}_{n+1} - Q^{-2}b^{*}_{n+1}b^{*}_{n+2},
\label{GOYmod}
\end{equation}
where the complex variable $b_{n}=k_{n}u_{n}$ should be understood as the
Fourier component of the gradient velocity field at wavenumber $k_{n}=Q^{n}$
and the integer $n$ runs from $0$ to $+\infty$. Throughout this paper,
usual values of parameters $\epsilon=0.5$ and $Q=2$ will be assumed. 
This choice yields values of scaling exponents $\zeta_{p}$ of the statistical
moments $\langle |u_{n}|^{p}\rangle$ very close to those predicted by the She-L\'ev\^eque formula\cite{SL94}~:
\begin{equation}
\zeta_{p}=h_{0}p +d_{0}[1-{\beta }^{p}]\,,
\label{SheLev}
\end{equation}
with $h_{0}=1/9$, $d_{0}=2$, and $\beta=(2/3)^{1/3}$. Equations 
(\ref{GOYmod}) admit formally self-similar solutions of the type :
\begin{equation}
b_{n}(t)=Q^{nz} g(v=Q^{nz}(t-t^{*}))\,.
\label{selfsim_1}
\end{equation}
Provided the scaling function $g$ vanishes for $v \rightarrow -\infty$ 
and goes to a constant for $v \rightarrow 0$, (\ref{selfsim_1}) describes
nothing but an invading front reaching the smallest
scales in finite time and leaving in its trail a spectrum of slope $z$ in
logarithmic scale. In the $b_{n}$-representation (the most relevant,
dynamically), such a front appears
like a soliton-like localized structure. For $z=2/3$, Kolmogorov scaling
is recovered.

\begin{figure}[p]
\centerline{\psfig{file=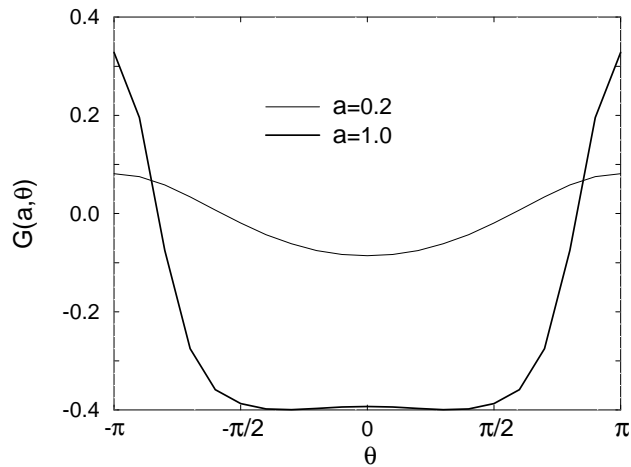,height=9cm}}
\caption{The logarithmic gain $G(a,\theta)$ versus $\theta$ for $a=0.2$ and $a=1.0$.}
\label{Fig.1}
\end{figure}

\begin{figure}[p]
\centerline{\psfig{file=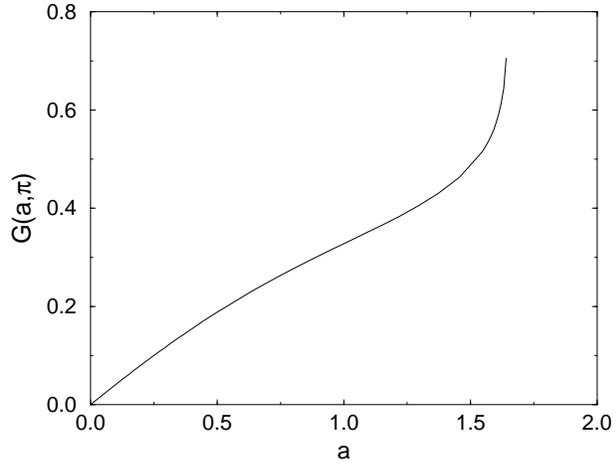,height=9cm} }
\caption{The optimal gain $G(a,\pi)$ versus the amplitude of the collision.
 }
\label{Fig.2}
\end{figure}

\begin{figure}[p]
\centerline{\psfig{file=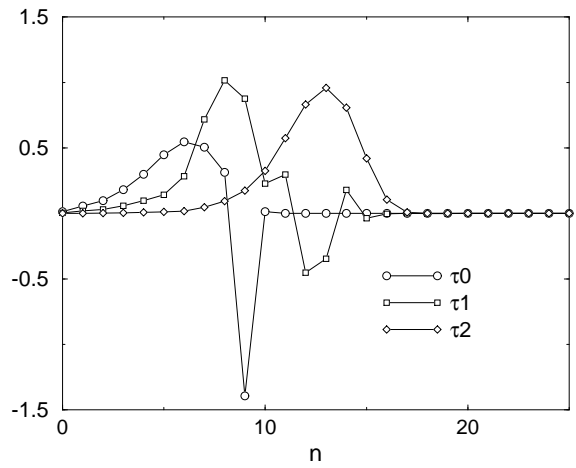,height=9cm} }
\caption{Evolution of the normalized field $C_n(\tau)=\frac{\vect{b}_n}{\protect\sqrt{(\vect{b},\vect{b})}}$ during a collision of amplitude $a=1.5$, close to the pulse-spliting threshold.}
\label{Fig.3}
\end{figure}

It is convenient to build from the $b_n$'s an infinite-dimensional vector 
$\vect{b}$,
whose squared norm $(\vect{b},\vect{b})=\sum_{n=0}^{\infty} 
|b_{n}|^{2}$ would be the enstrophy in real flow. The trend towards
enstrophy blow-up can  be cured by concentrating on the
dynamics of the unit vector $\vect{C}=\frac{\vect{b}}
{\sqrt{(\vect{b},\vect{b})}}$ and working with a desingularizing time
variable $\tau $ such that 
$\frac{d\tau }{dt} =\sqrt{(\vect{b},\vect{b})}$.
By doing so, it was observed in 
\cite{DG97} that every initial condition of finite enstrophy leads to the
same asymptotic state, up to a time translation, of the form 
$b_{n}(\tau) = e^{A_{0}\tau } \Phi_{0}(n-v_{0}\tau)$.
 Further, the envelope function $\Phi_{0}$ turns out to be purely real and positive up to trivial phase symmetries of the model \cite{BBP} leaving
 unaffected the energy flux $\epsilon_n$ from shell to shell defined as~:
\begin{equation}
\label{def:flux} 
\epsilon _n= Q^{-2n}\,\Re\left[ (1-\epsilon) b_{n-2} b_{n-1} b_{n} + b_{n-1} b_{n}b_{n+1}\right]\,.
\end{equation}
From the easily measurable values of the velocity $v_{0}$ and the growth rate $A_{0}$, the estimate $z_{0}=\frac{A_{0}}{v_{0}\log Q}= 0.72$ could be deduced. This is
 much lower than the exponent of the most intense event observed
in full simulations of the GOY model, which, according to (\ref{SheLev}), 
should be $z_{max}=1-h_{0}= 0.89$.\\
To bring out the rules of interaction of such coherent pulses with a
turbulent background, we first let one of them collide with some
localized activity residing on a shell downstream. The system was
prepared in a state consisting of a well-formed pulse
(of unit norm and real positive) and, three shells in front of its center,
 a ``defect'' of amplitude $a$ and relative phase $\theta $ 
corresponding to an initial perturbation of the form
$\delta b_{n}=a\exp(i\theta) \delta_{n,n_{0}}$. 
The dynamical rescaling method outlined before makes it very easy
to compare the amplitudes of the pulse, with or without collision,
when it crosses a shell far beyond the place of the collision. 
The relevant information is contained in the logarithmic amplitude gain
$G(a,\theta )$ defined as~:
\begin{equation}
\label{def:Gain}
G(a,\theta)= \lim _{n\gg n_{0}} \log \frac{|b_n^{coll}|}{|b_n^{0}|}\,.
\end{equation}
Fig.~\ref{Fig.1} shows the variation of $G$ with $\theta $ for two
values of the amplitude, whose first one, $a=0.2$, is still in the linear response regime. 
The first interesting observation is that a strong phase mismatch is
 required to enhance the growth of the pulse; the higher the strength of the
 collision, the more stringent gets this condition. 
Fig.~ \ref{Fig.2} shows the behaviour of $G$ as $a$ increases, in the
optimal case $\theta =\pi $. What ultimately limits the growth of the
incident pulse is the fact that the new structure forming in front of it, as 
a result of the collision, gets too fast and starts to lead its own life. By
 contrast, we
see in Fig.~\ref{Fig.3} how for $a=1.5$, i.e. slightly below the stability threshold, the latter is finally caught up by the former.
  
In order to induce corrections in scaling exponent $z$,
such collisions should occur repeatedly all along the cascade. It is
 reasonable to think that collisions with ``defects'' of large amplitude will
keep their high efficiency only if they are sparse enough, because they 
involve long-lived intermediate states presumably prone to splitting 
instabilities. To get a better feeling for the order of magnitude of exponent
corrections $\Delta z$ one may expect from such a mechanism, we just
divided $G(a,\pi)$ by the number $\Delta n(a)$ of steps the original pulse 
has to go forward before recovering its initial shape. 
This conservative estimate yields corrections in scaling exponents which 
can get as high as 0.16 or 0.18 on the $\Delta n =2$ and $\Delta n =3$ branches, corresponding respectively to $0.2 < a<0.6$ and 
$0.6 <a<1.2$. Those are the orders of magnitude expected to bridge the gap between $z_0$ and $1-h_0$.\\ 

\begin{figure}[p]
\centerline{\psfig{file=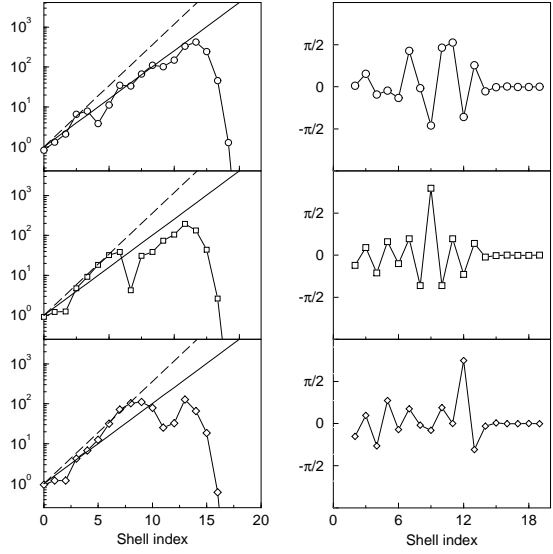,height=9cm}}
\caption{Three succesive snapshots of the propagation of a pulse for 
$Re=10^{6}$. Amplitudes $b_n$ are depicted on the left side, together with power laws $Q^{nz}$ for $z=2/3$ (solid line) and $z=0.85$ (dashed line), while
phases $\Psi _n$ are shown on the right. }
\label{Fig.4}
\end{figure}

The analysis of true signals corroborates to a large extent the two-fluid
picture we propose. In order to detect coherent structures, we just 
selected series of local maxima of $\epsilon _n$, starting from the top of the cascade and
going downwards, such that at each step $|b_n|$ grows by a factor larger than some prescribed value,
 conveniently written as $Q^{z}$. By letting the effective exponent $z$ vary,
 one scans events of various singular strength. Furthermore, coherency is
 controlled by demanding that the time interval $t_{n+1}-t_{n}$ between
 the occurence of local extrema on neighbouring shells never exceeds the
local turn-over time $\sim Q^{-nz}$ compatible with the scaling of
the singular event under examination. Figure \ref{Fig.4} depicts
such an event at three successive times for a Reynolds number 
$Re=10^{6}$. The imposed growth factor was $Q^{0.85}$, which
this particular pulse failed to achieve well before reaching the
dissipative shell of index $n_{d}=15$. 
The left side of Fig.~\ref{Fig.4}, showing $|b_n|$ versus $n$, 
reveals that the coherent structure
emerges from a disorganized K41-ramp, whose level does not seem to vary
 significantly during the time of observation (the length of the cascade is
however too short to make a definitive statement on this last point,
whose importance will be stressed later on).
On the right side of the same figure are shown the corresponding patterns of 
the gauge invariant phase $\Psi _n=\theta _{n-2}+\theta _{n-1}+\theta_{n}$ entering into the expression (\ref{def:flux}) of $\epsilon _n$. 
One notes the reduction of the amplitude
of spatial oscillations of the phase in the trail of the coherent structure,
as well as the presence of a well-established phase defect just in front of 
it, able to induce a local change of sign in the energy flux. The same
features were observed for every singular event we
analyzed in this way. Altogether, they fit in nicely with the conclusions
drawn from the study of elementary collisions. They also 
strongly suggest that the random background, necessary to feed 
anomalous growth of the coherent component, may simply obey
mean-field Kolmogorov scaling.\\

\begin{figure}[p]
\centerline{\psfig{file=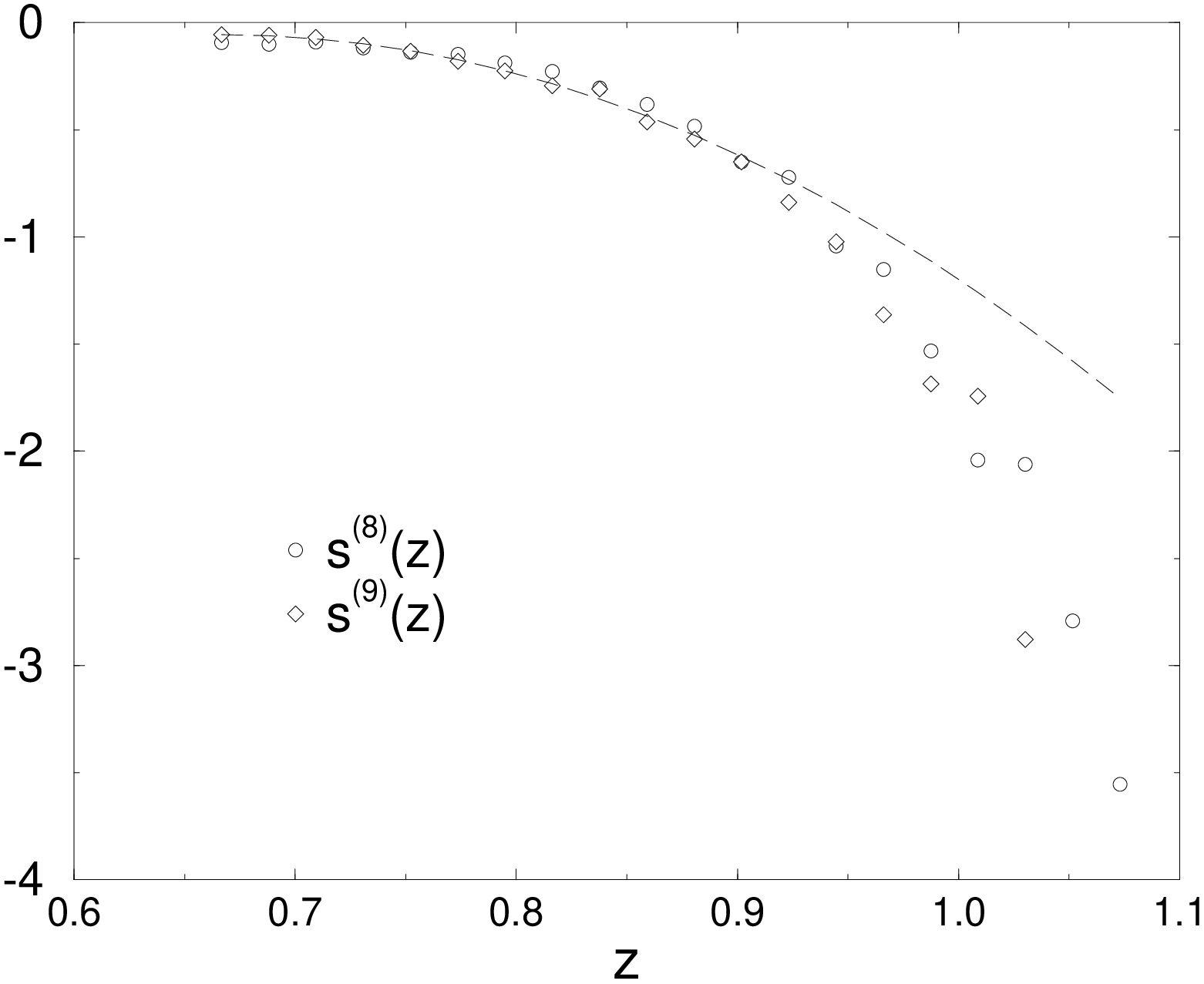,height=9cm}}
\caption{Estimates for the Cram\'er's function $s(z)$, extracted from
pdf's $P_{n}(z)$. Statistics was run over $3\times 10^{4}$ turn-over times of
 large scales.}
\label{Fig.5}
\end{figure}

At this point we are facing an interesting dilemma. 
If incoherent fluctuations were obeying some intrinsic scaling fixed
 once for all, say of Kolmogorov-type, we would expect unifractal statistics in
 the limit of infinite Reynolds number because only the scaling exponent $z_0$
 would survive to the increase of the cascade length. 
Multiscaling properties of the GOY model can be preserved asymptotically
within our two-fluid picture,
if and only if the turbulent background manages by a way or the other to
stay at par with the coherent structure propagating into it. 
The simplest way to describe this situation is provided by the following
 stochastic dynamical system~:
\begin{equation}
\label{def:syststoch}
\frac{d\vect{b}}{dt}= \vect{N}[\vect{b}] + \vectg{\eta}\,,
\end{equation}
where $\vect{b}$ now embodies the coherent part of the velocity field,
 $\vect{N}[\vect{b}]$ is the nonlinear kernel of (\ref{GOYmod}) and 
$\vectg{\eta}$ is a Gaussian random force, delta-correlated in time,
whose correlations read~: 
\begin{equation}
\label{def:correl1}
\langle \eta _n^{*}(t) \eta_{n'}(t')\rangle =\Gamma \,(\vect{b},\vect{b})^{3/2}\delta _{nn'} \delta (t-t')\,.
\end{equation}
with $\Gamma $ small for the sake of consistency.\\
Within an adiabatic approximation keeping the shape of the 
soliton frozen, dynamics (\ref{def:syststoch}) reduces to 
the biased Brownian motion of two collective variables $n(\tau)$ and 
$B(\tau )$, respectively the position of the soliton and the logarithm of its 
amplitude ($\tau$ is the proper time introduced before). The density of probability $P_{n}(z)$ for developing an effective growth
 exponent $z$ after $n$ steps appears then as a sum over all random walks 
starting from the origin (i.e. $n(0)=B(0)=0$) and
such that $B(\tau )=nz\log Q$ at the time $\tau $ of their first visit to 
the $n^{th}$ shell. 
For $n\gg 1$, $P_n(z)$ takes the form expected within multifractal 
descriptions \cite{F95}~:
\begin{equation}
\label{Proba}
P_{n}(z) \sim \sqrt{n}\exp[n s(z)+s_1(z)],
\end{equation}
where the Cram\'er's function $s(z)$ reads~:
\begin{equation}
\label{def:S}
s(z)= \frac{c}{\Gamma } \left[(z/z_0)+a^2
-\sqrt{\left((z/z_0)^2+a^2\right)\left(1+a^2\right)}\right].
\end{equation}
In the above equation, $a^2=\frac{\left(\partial _{\tau}\vectg{\Phi}_0,\partial_{\tau}\vectg{\Phi}_0\right)}{{A_0}^2}$ and  $c=A_{0}z_{0}\log Q$ are two numbers related only to the properties of the ideal self-similar solution.
The bad thing with (\ref{def:S}) is that it predicts a linear decrease
of $s(z)$ at large $z$ which we believe to be an artefact of the 
approximation used to solve (\ref{def:syststoch}) and, in any case, is not confirmed by numerics.\\
We tried indeed to extract $s(z)$ from true signals, by investigating the statistics of local maxima of $\epsilon_n$ exceeding 1 
(a value of the order of the average energy flux). Scaling
 exponents $z$ were computed as $z=\frac{2}{3} +\frac{1}{3(n-2)} \log _{Q}
\left[\frac{\epsilon_{n}(t_{i}^{(n)})}{\epsilon_{2}(t_{j(i)}^{(2)})}\right]$,
where $t^{(n)}_{i}$ denotes the time of occurence of the $i^{th}$ maximum 
on the $n^{th}$ shell, and $j(i)$ is the greatest integer such that all shells
of index between 2 and $n$ present at least one local maximum within the
time interval $[t^{(2)}_{j(i)}, t^{(n)}_{i}]$. 
For $Re=10^{6}$, the
resulting pdf $P_{n-2}(z)$ was observed to vary with $n$ in a way 
compatible with (\ref{Proba}) only in a very narrow window of shell
indices $7 \leq n \leq 9 $. We plot in
Fig.~\ref{Fig.5} $s^{(n)}(z)=\log \sqrt{\frac{n-3}{n-2}}
\frac{P_{n-2}(z)}{P_{n-3}(z)}$ for $n=8$ and 9, which should give a fair
account of the asymptotic Cram\'er's function $s(z)$. 
It is worth noting that deviations from a parabolic shape 
are hardly perceptible for values of $z$ as high as $0.87$, i.e. not
very far from the maximal exponent postulated in She-L\'ev\^eque's approach.
However, there is no hint of an abrupt cut-off on $z$, in
contradiction with the conclusions of \cite{LS97}. It would be interesting
to see whether fully nonlinear treatments of 
the stochastic dynamical system (\ref{def:syststoch}), based on instanton techniques similar to those recently introduced in the context of turbulence
 \cite{IN}, yield a better agreement with Fig.~\ref{Fig.5} than
the analytical result (\ref{def:S}) does. Work in this direction is
 in progress.\\

To conclude, let us emphasize that our theory, though still calling for improvement, rests on robust physical ingredients which could well have their
 counterparts in
Navier-Stokes dynamics. The renormalization of coherent structures by their environment turns out to be so strong
 that the existence of ideal self-similar objects, leading to finite
time blow-up of the enstrophy in the zero-viscosity limit,  
is certainly not a prerequisite for the whole approach to make sense. 
It remains to see how the random background should be modelled in the
context of Navier-Stokes equations.

%
%
%
%

%
%

\end{document}